\documentclass[pop,fleqn]{w-art}
\usepackage{times}
\usepackage{w-thm}
\usepackage{graphicx}
\begin{document}
\keywords{Quantum computation, control theory, segmented ion trap}
\subjclass[pacs]{02.30.Yy, 32.80.Pj,03.67.Lx}
\title{Optimization of segmented linear Paul traps and transport of stored particles}
\author{Stephan Schulz, Ulrich Poschinger, Kilian Singer\footnote{Corresponding author: \textsf{kilian.singer@uni-ulm.de}}
 and Ferdinand Schmidt-Kaler}
\address{Abteilung Quanten-Informationsverarbeitung,  Universit\"at Ulm, Albert-Einstein-Allee 11, D-89069 Ulm, Germany}

\begin{abstract}
Single ions held in linear Paul traps are promising candidates for a
future quantum computer. Here, we discuss a two-layer
microstructured segmented linear ion trap. The radial and axial
potentials are obtained from numeric field simulations and the
geometry of the trap is optimized. As the trap electrodes are
segmented in the axial direction, the trap allows the transport of
ions between different spatial regions. Starting with realistic
numerically obtained axial potentials, we optimize the transport of
an ion such that the motional degrees of freedom are not excited,
even though the transport speed far exceeds the adiabatic regime. In
our optimization we achieve a transport within roughly two
oscillation periods in the axial trap potential compared to typical
adiabatic transports that take of the order 10$^2$ oscillations.
Furthermore heating due to quantum mechanical effects is estimated
and suppression strategies are proposed.
\end{abstract}

\maketitle
\renewcommand{\leftmark}
{Optimization of segmented linear Paul traps and transport of stored
particles}

\tableofcontents  
\section{Introduction}
\label{sec:intro}

With a series of spectacular experiments the ion trap based
quantum computing has proven its prominent position for a future
quantum computer among the list of candidates \cite{ROADMAPS}.
Starting with two-qubit gate operations \cite{SCHM03a,LEIB2003},
long lived two-qubit entanglement
\cite{ROOS2004,HAEFFNER2005,LANGER2005}, teleportation experiments
\cite{RIEBE2004,BARRETT2004}, and different sorts of multi-qubit
entangled states \cite{SACKETT00,HAL2005,ROOS2004,LEIB2004}, the
record for qubit-entanglement is currently presented in a 6-qubit
cat state and a 8-qubit W-state \cite{LEIB2006,HAFF2006}. Future
improvement is expected using the technique of segmented linear
Paul traps which allow to shuttle ions from a ``processor'' unit
to a ``memory'' section \cite{KIP2002}. In such a quantum
computer, strategies of quantum error correction will be critical
for the successful operation. However, as a result, many
additional ancilla qubits are required and a large fraction of the
computational time will be consumed by shuttling ions between
different segments. Detailed simulations \cite{CHUANG2006} show
that as much as 99$\%$ of the operating time would be spent with
the transportation processes. The time required for the transport
should be reduced such that the gate times are improved and
decoherence processes are reduced.

Thus, we assume that the improvement of these transport processes
is necessary. In recent experiments \cite{LEIB2006,ROWE2002}, the
shuttling has been carried out within the adiabatic limit, such
that the time required for the transport by far exceeds the
oscillation time of the ion in the potential. It is a common
misbelief that this adiabatic transport is necessary to avoid the
excitation of vibrational quanta. In this spirit, we investigate
in this paper the optimization of fast and non-adiabatic
transportation by applying classical optimal control theory. Our
simulations allow to predict the time sequence of control voltages
such that ion heating is suppressed.

Certainly, non-optimized fast transport of qubit ions into the
processor unit followed by sympathetic cooling of a different ion
species \cite{BARRETT03,DREWSWEN2004} would be an alternative
strategy. However, the necessary cooling time would render the
overall computational time even slower. First experiments show that
the qubit coherence is maintained during a transport, but that the
vibrational quantum state may typically not be well conserved after
a fast shuttle of the ions. This impedes further qubit operations.

In the first section of this paper we start by numeric calculations
of the electric trapping potential for ions and show how to optimize
the geometry of a two-layer microstructured segmented trap
\cite{LEIBFRIED2004,STICK2006}. The same techniques may be applied
for the optimization of planar ion traps
\cite{CHIA2005,MADSEN2004,PEARSON2006,SEIDE2006}. In the
second section we optimize the transport of a single ion between two
regions and illustrate the application of optimal control theory
\cite{MABUCHI2005}. Even though shuttling is fast, we can show
that an optimized non-adiabatic transport does not lead to
significant heating.

\section{Optimization of a two-layer microstructured ion trap}
\label{sec:field}

\begin{figure}[t]
\centering
\begin{center}
\includegraphics[width=.97\textwidth]{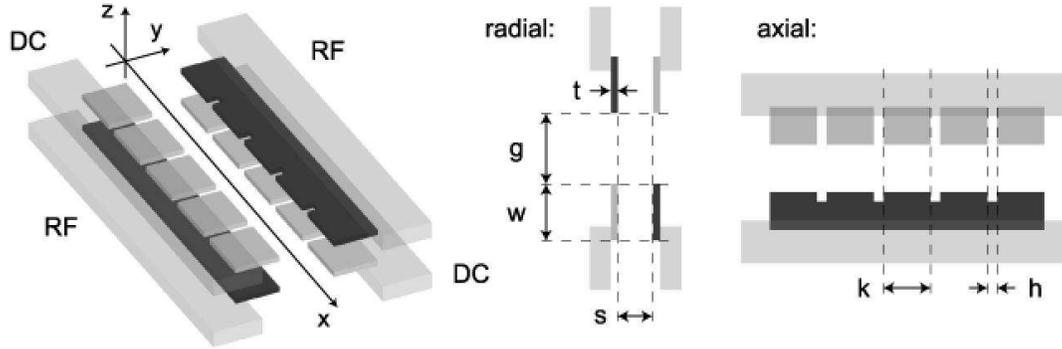}
\end{center}
\small \caption{Scheme of a two-layer micro structured segmented
linear trap: The two electrode layers have a thickness $t$ and are
separated by the distance $s$. The length of the trapping electrodes
is $w$, the radio frequency electrodes (RF) and the segmented
electrodes (DC) are separated on each layer by the gap $g$ (radial
direction). The RF voltage is applied on two continuous electrodes
(black) and the static voltages are applied on the segmented DC
electrodes (gray). The DC electrode segments have the length $k$ and
are separated by a gap $h$. The symmetry axis is later denoted as
the x- or axial direction.} \label{fig:schemafalle}
\end{figure}

\begin{figure}[b]
\centering
\begin{minipage}{0.5\linewidth}
\begin{center}
\includegraphics[width=.75\textwidth]{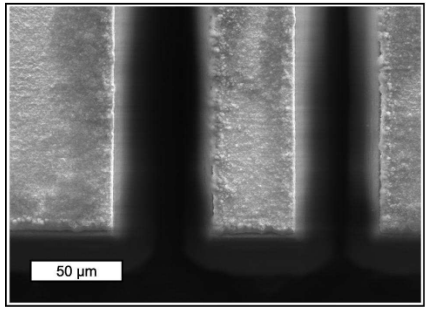}
\end{center}
\end{minipage}
\begin{minipage}{0.45\linewidth}
\caption{Detail view on the trap chip, 5$\mu$m gold plated alumina
of 125$\mu$m width cut by a fs-pulsed Ti:Sapphire laser. The
scanning electron microscope picture shows several DC electrode
segments of a single layer.} \label{fig:2}
\end{minipage}
\end{figure}

The idea of segmented linear Paul traps has been proposed to
realize a scalable quantum computer
\cite{CIR95,WINELAND1998,KIP2002}. Typically, these trap structures are fabricated out of etched semiconductor structures \cite{STICK2006} or gold plated insulators structured by microfabrication techniques \cite{HENS2006}. Segmented traps come in various shapes and can be categorized by the number of electrode layers
forming the trapping potential:
Planar traps with one layer only, two-layer traps that are
composed of two microstructured planar chips and traps with a higher
number of electrode layers. In our discussion, we will focus on the
two-layer geometry which is shown in Fig.~\ref{fig:schemafalle}.
To illustrate the methods of fabrication, Fig.~\ref{fig:2} gives
an SEM picture of a gold plated laser cut alumina
wafer. Here, with fs-laser ablation \cite{CHICHKOV96}, the cuts
are clean and show a spatial resolution of about 2$\mu$m. The
DC-electrodes may be cut in form of ``fingers'' to reduce the
insulating surface seen directly from the ion position. This
reduces the influence of the possibly charged surfaces to the trap
potential and has also been shown to reduce heating effects of the
ion motion. Two structured wafers are assembled to form a two-layer trap
geometry as shown in Fig.~\ref{fig:schemafalle}.

\subsection{Design objectives}

What are the optimal dimensions and aspect ratios in such an ion
trap structure? What are the optimal electric trap parameters?

\paragraph{Radial configuration}
At first, we aim for a high secular trap frequency
$\omega_{\textrm{sec}}/2\pi$, such that there is a tight dynamical
confinement of the ions within the Lamb-Dicke regime. The
confinement should typically reach frequencies of several MHz in the
radial direction. The required radial frequencies should be achieved
with moderate voltages on the electrodes of several hundreds volts.
Therefore, the RF trap drive may not exceed the break-through
voltage - a limitation which plays a significant role in the case of
very small traps \cite{STICK2006,CHIA2005}.

A second aspect is the anharmonicity of the radial trapping
potential. From the fact that linear traps with optimized electrode
shapes have been shown to load large crystals of ions
\cite{DREWSEN2006}, we would try to improve the loading rate by
reducing non-harmonic contributions to the potential. Especially for
larger q-values when the trap drive power is chosen relatively high,
non-linear resonances have been observed \cite{WERTH96}. This
confirms that even small anharmonicities are relevant in the case of
large crystals.

\paragraph{Axial configuration}
In order to maintain the linear appearance of the ion crystals, the
axial trap frequencies have to be lower than the radial frequency.
Nevertheless, the axial frequencies $\omega_{\textrm{ax}}/2\pi$ should exceed a
few MHz. Then, cooling techniques are simpler \cite{ESCHNER2003},
gate operations may be driven faster, and a faster adiabatic
transport of ions between segments may be achieved. Ion transport
between axial segments requires a fast temporal change of the trap
control voltages on the order of several $\mu s$. This is accomplished
by controlling the DC-electrode voltages by means of fast digital-to-analog converters
(DAC) and would be technically much more involved for high
voltages.\footnote{Therefore the geometry should also take into account the
limited voltage range of the DACs.}

Furthermore, single ions will have to be split off and merged to ion
strings throughout the operation of a segmented ion trap quantum
computer. The investigation of splitting and merging operations is
not within the scope of this paper \cite{POSCHINGER2006}, however,
it was pointed out that a highly non-harmonic axial potential improves this
situation \cite{HOME2004}; it implies certain geometrical
ratios in the axial trap construction.

\begin{figure}[t,b]
\centering
\begin{minipage}{0.55\linewidth}
\begin{center}
\includegraphics[width=0.95\textwidth]{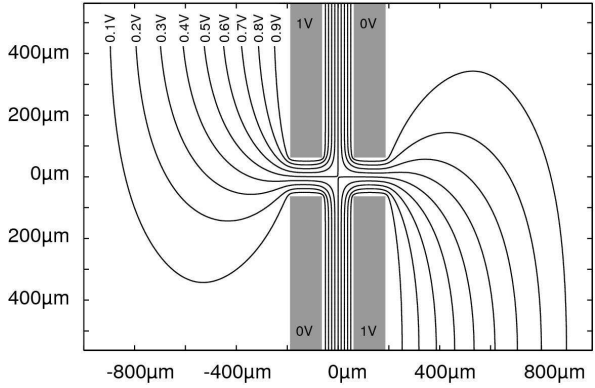}
\end{center}
\end{minipage}
\begin{minipage}{0.4\linewidth}
\caption{Electric potential of the two-layer microstructure trap in radial
direction ($yz$ cross section). Ions are confined by a
pseudo-potential on the $x$-axis. Here, the potential lines are
normalized to a trap drive amplitude $U_{\textrm{rf}}$ of 1V. In the
central trapping region near the x-axis the electric potential may be approximated by a
quadrupole potential as the radial harmonic pseudopotential for single ions.} \label{fig:3}
\end{minipage}
\end{figure}

\begin{figure}[b]
\centering
\includegraphics[width=.95\textwidth]{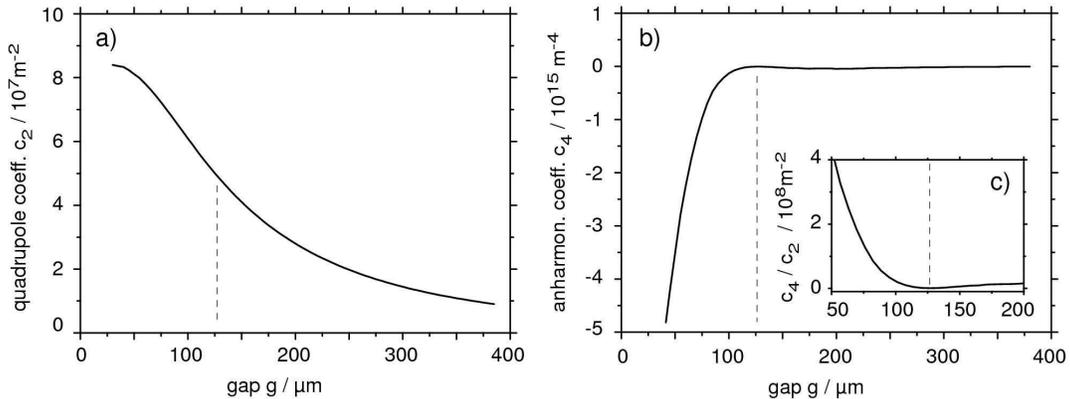}
\caption{a) Numerical simulation of the trap stiffness $c_2$ in
radial direction as a function of the slit width~$g$. For a slit width
of 126$\mu$m  we estimate a value of 5.5$\cdot$10$^7m^{-2}$ which
corresponds to a radial frequency of $\omega_{\textrm{rad}}/2\pi=$~5MHz.
b) Dependence of the fourth order parameter $c_4$ to the slit width $g$.
The dashed line near 126$\mu$m indicates the optimization result.
c) The normalized hexapole coefficient $c_4$ normalized by the
quadrupole coefficient $c_2$ indicates the descreased loss of
the trap drive power.} \label{fig:4}
\end{figure}
\subsection{Operating mode and modeling of the segmented linear Paul trap}

Linear Paul traps are characterized by a two-dimensional dynamical
confinement in the radial direction ($yz$ plane) and a static
confinement in the axial direction ($x$-axis). The applied radio
frequency $\omega_{\textrm{rf}}/2\pi$ to the RF electrodes (see
Fig.\ref{fig:schemafalle}) generates a dynamical electric potential
$\phi_{\textrm{rad}}(y,z,t)$ which leads to a strong confinement of
single ions along the axial direction at the radio frequency node.
Typically, the axial potential $\phi_{\textrm{ax}}(x,t)$ formed by
the quasistatic voltages applied to the segmented DC electrodes is
weaker than the radial confinement to support a robust alignment of
the linear ion string. The shape of this axial potential depends on
the geometry of the segmented DC electrodes. The time-dependent
variation of the DC control voltages allows to transport
ions in the axial direction without micromotion. We separate the
numerical optimization of the linear Paul trap into a radial and
axial calculation - first, the radial geometry configuration is
optimized for strong confinement in the RF node, then the axial
electrode geometry is calculated based on the radial geometry.

The lowest-order approximation of the dynamical trap potential
$\phi_{\textrm{rad}}(y,z,t)$ in radial direction is similar to that of a
quadrupole mass filter \cite{Paul1958}. The geometric factor $c_2$
describes the quadrupole potential strength in both radial
directions for a symmetric radial electrode geometry:
\begin{equation}
\phi_{\textrm{rad}}(y,z,t)=c_2/2\:(y^2-z^2)\:(U_{\textrm{dc}}+U_{\textrm{rf}}\cdot
\cos(\omega_{\textrm{rf}}\:t)) \label{pot1st}
\end{equation}

An ion trajectory is described as a superposition of a harmonic
secular motion at frequency $\omega_{\textrm{sec}}=
\omega_{\textrm{rf}}/2~\sqrt{a+q^2/2}$ (lowest order approximation)
and the superposed micromotion at the radio frequency
$\omega_{\textrm{rf}}$. The frequency of the secular motion is
characterized by the dimensionless stability
parameters $a$ and $q$ of the radial motion \cite{Leibfried2003}
which depends on mass $m$ and charge $e$ of the ion, the RF
amplitude $U_{\textrm{rf}}$ applied to the RF electrodes and the
static voltages $U_{\textrm{dc}}$ applied to the segmented electrodes of the
trap:

\begin{equation}
a=\frac{4 e\:U_{\textrm{dc}}}{m\:\omega_{\textrm{rf}}^2} \; c_2 \quad
, \qquad q=\frac{2 e\:U_{\textrm{rf}}}{m\:\omega_{\textrm{rf}}^2}  \; c_2 \label{aq}
\end{equation}

A two-dimensional domain of the stability parameters $a$ and
$q$ defines a region of stable trajectories as solutions of the
classical equations of motion\footnote{We discuss the optimization
in the so-called lowest stability region including $a=0$ and $q \leq
0.9$}. In general, the electrode configurations result in an
electric potential that may be expanded in spherical multipole
components, where the quadrupole contribution $c_2$ represents the
dominating part for reasonable Paul trap geometry; the hexapole
contribution $c_4$ contributes mainly to the non-harmonic part.

The quadrupole approximation of the confining potential is
inaccurate if the electrode shapes deviate strongly from the ideal
hyperbolic form. As a result, anharmonicities and coupling terms
appear inside the stability region \cite{WERTH96}. As the radio
frequency voltage is portioned to various higher order terms and
not only to the quadrupole contribution of the potential a loss of
the trap stiffness c$_2$ is observed (Fig. 4). For simplicity we idealize
$U_{\textrm{dc}}$ as zero and characterize the anharmonicity of the
pseudopotential in radial direction along the two radial
principal axes, here denoted by a radial coordinate $r(y,z)$, by the
leading terms of the following polynomial expansion:

\begin{eqnarray}
\phi_{\textrm{rad}}(r(y,z),t) \propto  \hspace{1mm} \sum_{n}{c_n \hspace{1mm} r^n}
\label{potexp}
\end{eqnarray}

Because of the radial electrode symmetry the odd-numbered terms
$c_1, c_3, \ldots$ are negligible and the potential offset $c_0$ is
irrelevant. The optimization of the radial trap potential leads to a
suppression of the higher order potential contribution, such that
the hexapole term $c_4$ as the leading non-harmonic contribution is
small.

Based on the geometry for an optimized radial confinement the
axial static trap potential along the symmetry axis $x$ can be
analogously expanded,
\begin{eqnarray}
\phi_{\textrm{x}}(x,t) \propto \sum_{n}{d_n \hspace{1mm} x^n}.
\label{potexp}
\end{eqnarray}
The axial potential properties are determined by the segmented
electrode geometry, especially the axial width of the individual
electrode segments. An optimal axial confinement of the ion
requires a maximum quadratic term $d_2$. The transport of a single
ion between axial segments is facilitated if the potentials from
adjacent segments exhibit a large overlap. For the splitting
operation of an aligned two-ion crystal into single ions in
independent axial potentials, Steane {\itshape et al.}
\cite{HOME2004} suggest a potential shape with a maximum quartic
term $d_4$ and minor quadratic contribution $d_2$.

\begin{table}[htbp]
\begin{center}
\begin{minipage}{15cm}
\renewcommand{\footnoterule}{}
\begin{center}
\begin{tabular}{|l|c|c|c|c|c|c|c|c|c|c|} \hline
&& $R$ &$c_2$ & $\omega_{\textrm{rf}}/2\pi$ & $U_{\textrm{rf}}$ &
$q$ & $\omega_{\textrm{rad}}/2\pi$ & $\omega_{\textrm{ax}}/2\pi$ &
$\Delta$
\rule{0in}{3ex} \\
&& [$\mu$m] &[$1/m^2$] & [MHz] & [V] &  & [MHz]& [MHz]&
[meV]\rule{0in}{3ex}\\[1ex] \hline \hline
Aarhus\,\cite{DREWSEN1998}& $^{24}$Mg$^+$ & $1750$ & $1.6\!\cdot\!10^5$ & $4.2$ & $2\!\cdot\!50...150$ & $0.2...0.6$ &
$0.3...0.8$ & $\leq\!0.4$  & $\leq\!10^5$\rule{0in}{3ex}\\
Innsbruck\,\cite{FSK2003}& $^{40}$Ca$^+$ & $800$ & $3.9\!\cdot\!10^6$ & $23.5$ & $700$ &
$0.6$ &
$5.0$ & $1.0$ & $10^3$\rule{0in}{3ex}\\
Michigan\,\cite{HENSINGER2006}& $^{112}$Cd$^+$ & $100$ & $2.2\!\cdot\!10^7$ & $48.0$ & &
$0.3$ &
$5.0$ & $2.5$ & \rule{0in}{3ex}\\
Simulation& $^{40}$Ca$^+$ & $89$ & $5.3\!\cdot\!10^7$ & $50.0$ & $120$ & $0.3$ &
$5.0$ & $2.5$ & $300$\rule{0in}{3ex}\\
Michigan\,\cite{DSTICK2006}& $^{112}$Cd$^+$ & $30$ & $4.7\!\cdot\!10^8$ & $15.9$ & $8$ & $0.6$
&
$4.3$ & $1.0$ & $80$ \rule{0in}{3ex}\\ \hline
\end{tabular}
\caption[trapdesign]{Trap design parameters of several types of
linear ion traps: The geometric trap size $R$ given by the minimal distance
between ion position and electrode surface, the quadratic
geometry factor $c_2$ of the radial cross section describes the
magnitude of the radial confinement at the given electrode
voltage, the trap drive frequency $\omega_{\textrm{rf}}/2\pi$
together with the trap drive voltage $U_{\textrm{rf}}$ and the RF
stability parameter $q$ results in the radial motional frequency
$\omega_{\textrm{rad}}/2\pi$. For comparison the axial motional
frequency $\omega_{\textrm{ax}}/2\pi$ is shown. The trap depth
$\Delta$ summarizes the confinement of a single ion.}
\end{center}
\end{minipage}
\end{center}
\end{table}

Relevant parameters of various linear ion traps are summarized in
Table 1. The Aarhus hexapole design with endcaps \cite{DREWSEN1998}
and the Innsbruck blade design \cite{FSK2003} show a traditional
macroscopic approach of mm-size linear trap design without
segmentation of the control electrodes. The Michigan trap designs,
the microstructured three-layer trap \cite{HENSINGER2006} and the
semiconductor two-layer trap \cite{DSTICK2006}, represent the
progress in the miniaturization of linear ion traps and the
segmentation of the control electrodes for the transport of single
ions and the splitting of ion crystals.

\subsection{Optimization of the radial potential}
In the first step we optimize the radial confinement of the trap.
The width of the slit is varied and the electric potential is
calculated, see Fig.~\ref{fig:3}. The distance of the two layers
is fixed to the thickness of a commercial alumina wafer
(125$\mu$m) which acts as a spacer. Then a variable parameter is
the width $g$ of the lateral laser cut in the trap chips, respectively
the distance between the RF and the DC-electrodes of the trap chips. We find
that the radial confinement increases with decreasing slit width $g$,
see Fig.~\ref{fig:4}. Interestingly, for this geometry, the radial
potential is almost harmonic since the fourth order parameter
$c_4$ is nearly vanishing. For a width of 126$\mu$m the radial frequency
of $\omega_{\textrm{rad}}/2\pi= $~5~MHz is reached for a singly charged $^{40}Ca^+$ ion
with a peak RF voltage of 120~$V$ at 50~MHz.

\begin{figure}[t,b]
\centering
\includegraphics[width=.65\textwidth]{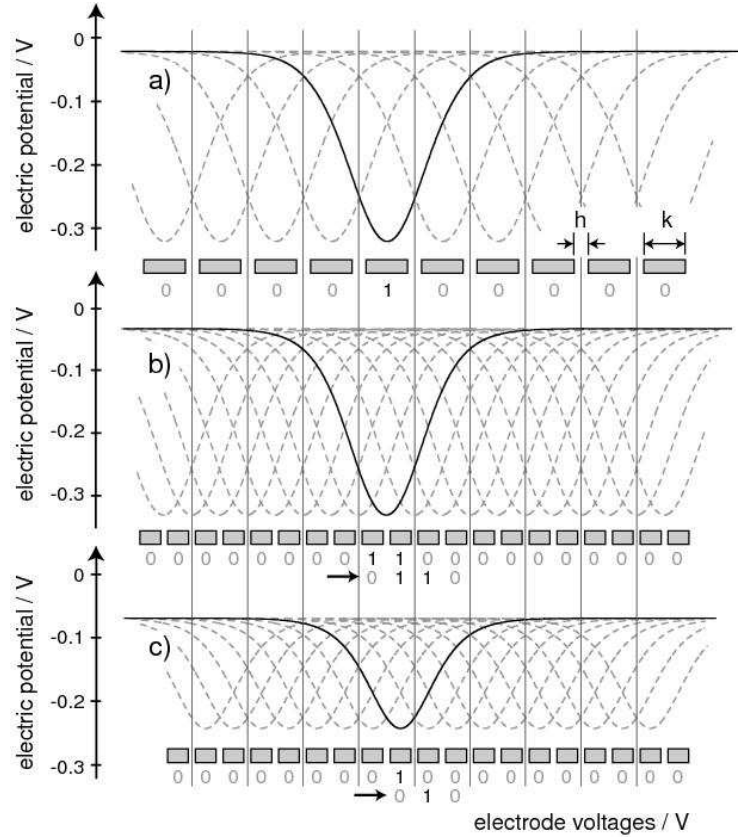}
\caption{Numerical calculation of the axial electric potentials
$\phi_{\textrm{ax}}$, all cases (a) to (c) for the optimized radial
slit width of $g=126~\mu$m. a) DC segments with a width of
$k=90~\mu$m and gaps between axial electrodes of $h=30~\mu$m result
in a maximal $d_2$ coefficient. The potential of the adjacent
electrode is plotted and shows only partial overlap (dashed gray).
b) Optimized transport scheme: The DC electrodes are divided into
equal parts with $k=$45$\mu$m and $h=$15$\mu$m. If both electrodes are at
the same voltage of 1V, the potential is nearly identical to the
optimized case (a). For transporting an ion, the potential minimum
is shifted by changing the voltages from ..0/0/1/1/0/0..
$\rightarrow$ ..0/0/0/1/1/0.. continuously.  The axial potentials
exhibit a large overlap which improves the transport of the ions. c)
Simplified transport scheme: Axial potential for DC segments with
$k=45~\mu$m and $h=15~\mu$m. Now, only a single segment is at 1~V.
For transporting the voltages are changed from ..0/1/0/0/..
$\rightarrow$ ..0/0/1/0/... } \label{fig:potfuertransport}
\end{figure}

\subsection{Optimization of the axial potential}
The optimization of the axial potential determines the
performance of the fast ion transport. Additional
requirements are a deep axial potential even with moderate DC control
voltages and the capability of the segments to split a two-ion crystal
into two single ions trapped independently in distinct potential
minima.

\begin{figure}[t,b]
\centering
\includegraphics[width=.95\textwidth]{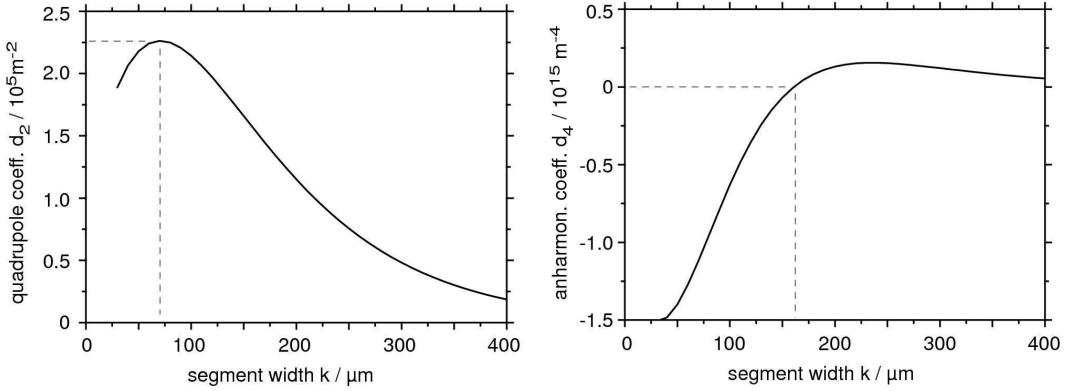}
\caption{a) Numerical simulation of the axial trap stiffness $d_2$
as a function of the segment width $k$ with $h=$30$\mu$m. b)
Dependence of the fourth order parameter $d_4$, the dashed line
near $k=$160$\mu$m indicates the zero crossing.}
\label{fig:dieletzte}
\end{figure}

In a first step we investigate the maximization of the axial trap
frequency $\omega_{\textrm{ax}}$ as a function of the segment width $k$ and the cut
width $h$, see Fig.\ref{fig:schemafalle}. The numerical
three-dimensional electric potential simulation depicted in Fig.~\ref{fig:dieletzte} shows the expected result: For a large
width of the segments the potential is expected to be shallow,
and for a very short width the electric potential falls rapidly off
from the electrode tips, such that again a weak confinement is
found. The maximum axial trap frequency is reached for a segment
width $k_{opt} \simeq$ 70~$\mu$m, a gap between DC and RF electrode $g$= 126$\mu$m
and a cut width of $h$= 30$\mu$m. Changing the
size of the segment width by 50$\%$ results only in a 20$\%$
variation of $d_2$ which is easily compensated by the DC voltage.

The trapping of single ions and transport require a different
electrode configuration. For the transport problem, it is
important that the potentials generated by adjacent DC segments
exhibit a large spatial overlap (see
Fig.~\ref{fig:potfuertransport}). To achieve both ideal trapping
and transport conditions we split each electrode into two parts
(b). For trapping we bias two neighboring electrodes with an equal
voltage in order to obtain a larger ``effective'' electrode. Due
to the smaller segmentation a better overlap of the
individual potentials is provided during transport. As the ion is
displaced during the transport process we expect that the
anharmonic terms $d_4, d_5, \ldots$ of the potential will cause heating, see
Sect.~\ref{anharmonic dispersion}. Therefore we have determined
the optimal effective segment width such that the $d_4$ term is
minimized. The results are shown in Fig.~\ref{fig:dieletzte}.

In the following, we will use the simplified transport scheme in Fig.~\ref{fig:potfuertransport}~(c)
with one start and one target electrode only and investigate the necessary time dependence of both DC
segment control voltages.

\section{Open loop control of ion transport}
\label{sec:transport}

After the optimization of all geometric trap parameters we now focus
on the optimization of the time dependent trap control voltages
which are applied to the DC segments in order to transport the ion: Our goal
will be to decrease the time required for the transport far below
the limit of adiabaticity, such that the transport is finished
within a single oscillation period only, with the constraint to
avoid vibrational excitation. To a good approximation the radial ion
confinement does not influence the axial transport between two
segments as the ion is moving along the central RF-node with
negligible micromotion. Note, that our calculation takes into
account two axial segments but may be adapted to a larger number of
segments, see Fig.~~\ref{fig:potfuertransport}~(c). The potential
which we use for the optimization of the ion transport is the result
of a boundary element calculation, see appendix~\ref{Anhang BEM}. In
order to transport the ion, the potential minimum is shifted by
changing the DC control voltages $u_i(t)$. Intuitively, we estimate
that a smooth acceleration and a smooth deceleration of the ion is
advantageous. Searching for the precise shape of the segment control
non adiabatic heating due to fast transport has to be minimized.

\subsection{Non-adiabatic heating sources}
For a transport duration approaching the timescale given by the
axial trap frequency, the following non-adiabatic effects are
expected to occur:
\begin{enumerate}
\item
{\em Classical displacement error:} The ion cannot adiabatically
follow  the potential minimum and starts oscillating, such that it
possesses excess energy after the transport process. This behavior
may be understood classically. In the quantum picture it corresponds
to the buildup of a nonvanishing displacement $\alpha$ during
transport.
\item
{\em Wavepacket dispersion heating:} With a spatial extension of
about 10 to 20~nm, the undisplaced wavefunction hardly senses any
anharmonicity in an  electric potential that is generated by 50 to
100~$\mu$m sized electrode structures. However, during the transport
the wavepacket undergoes significant excursions of a few $\mu$m out
of the minimum of the potential. Here, exposed to higher anharmonic
$d_4$ contributions, the  shape of the ion wavepacket disperses
which results in vibrational excitation.
\item
{\em Parametric heating:} As the control voltages are changed, the
harmonic frequency $\omega$ of the instantaneous potential is
temporarily varying. If the width of the wavepacket can not follow
the variation of $\omega(t)$ adiabatically, parametric heating to
higher vibrational states will occur.
\end{enumerate}

\subsection{Overview of the applied optimization strategies}

In the following Sect.~\ref{sec:oct} we minimize the classical
displacement error by applying the optimal control method. For the
optimization of this entirely classical error source we need to
optimize the ion's classical trajectory such that a cost function
- weighting the ion's phase space displacement after the transport
- is minimized. The solution obtained by the optimal control
algorithm does not show considerable heating by wavepacket
dispersion. However, we find a significant contribution of
parametric heating. A first guess would be to include an
additional term in the cost function to prevent parametric
heating, involving time derivatives of the control fields.

As this approach fails due to implementational difficulties, we
suppress parametric heating by an appropriate initial guess which
keeps the trap frequency perfectly constant. This is achieved by a
variable transformation from $u_{1,2}(t)$ to new parameters that
allow to decouple the strength of the potential and its minimum
position. Starting now the optimal control method yields a
solution that reduces the displacement error. Since the control
parameters are only slightly modified by the optimization
algorithm, the parametric heating and also the wavepacket
dispersion heating are negligible.

We conclude that in our case, the choice of variables that
decouple the essential optimization parameters
\cite{CALARCO2004,DORNER2005} and a well suited initial guess
function are helpful and maybe critical for the success of the
optimal control method.

\subsection{The optimal control method}
\label{sec:oct} This section will give an introduction to optimal
control theory applied to single ion transport. We use the method
derived from a variational principle with unbounded controls and
fixed final time \cite{OCTBook}. We consider the dynamics of a
singly charged single $^{40}$Ca$^+$ ion confined in a segmented
linear Paul trap. We assume that the ion is laser cooled to its
motional ground state\footnote{The calculation is valid also for
thermal and coherent states with modest excitation.} pertaining to
the axial degree of freedom. Neglecting the radial motion, the
motional state of a trapped ion is classically represented by a
coordinate vector $\vec{\xi}(t)=\left( x,v \right)^T$ in a two
dimensional phase space.

The equation of motion under consideration of two uniform
electrode segments with arbitrary voltages applied on them, reads
\begin{equation}
\dot{\vec{\xi}}=\vec{a}(\vec{\xi},\{u_i\})=\left(\begin{array}{c}
v\\-\frac{1}{m}\sum_i\frac{\partial V_i(x)}{\partial
x}u_i(t)\end{array} \right). \label{EqOfMotion}
\end{equation}

Here, the index $i$ runs over the two electrodes and $V_i(x)$ are
the normalized electrostatic potentials at electrode $i$. Eq.
(\ref{EqOfMotion}) then holds for arbitrary electrode voltages due
to the linearity of the Laplace equation.\\
Our goal is now to find time-dependent control voltages $u_i(t)$
that move the ion from the center of electrode 1 to the center of
electrode 2. We desire to have the ion at rest after the transport
process. The performance of a given control field is judged by the
cost function
\begin{equation}
h(\vec{\xi}(t_f))=\alpha\;(x(t_f)-x_f)^2+\beta\;v(t_f)^2,
\label{CostAtFinalTime}
\end{equation}
which is a measure of the phase space displacement at the final time
$t_f$. $\alpha$ and $\beta$ weight the contributions relative to
each other. Taking Eq. (\ref{EqOfMotion}) as a constraint for all
times $t$, we obtain the cost functional

\begin{eqnarray}
J(\vec{\xi},\vec{\xi}_p,\{u_i\})=\int_{t_0}^{t_f}\frac{\partial
h}{\partial \vec{\xi}} \cdot
\dot{\vec{\xi}}&+&\vec{\xi}_p\cdot\left(\vec{a}(\vec{\xi},\{u_i\})-\dot{\vec{\xi}}\right)
dt
\end{eqnarray}
where we have introduced the costate vector $\xi_p=\left( x_p,v_p
\right)^T$ as a Lagrange multiplier in order to guarantee that the
optimization result obeys the equation of motion
Eq.~(\ref{EqOfMotion}). The time dependence of all variables has
been dropped in the notation. For an optimal control field, $\delta
J=0$ has to hold, therefore the variational derivatives with respect
to $\vec{\xi},\vec{\xi}_p$ and $\vec{u}$ have to vanish. The
derivative with respect to $\vec{\xi}_p$ restores the equations of
motion Eq.~(\ref{EqOfMotion}) for the state vector, the derivative
with respect to $\vec{\xi}$ yields equations of motion for the
costate vector:

\begin{eqnarray}
\dot{\vec{\xi}}_p&=&-\frac{\partial\vec{a}}{\partial\vec{\xi}}\cdot\vec{\xi}_p
\Rightarrow \nonumber \\
\dot{x}_p&=&v_p \frac{1}{m}\sum_i \frac{\partial^2 V_i(x)}{\partial
x^2}u_i \nonumber \\
\dot{v}_p&=&-x_p. \label{CoEqOfMotion}
\end{eqnarray}

Variation of $J$ with respect to the control field leads to an
additional algebraic equation:
\begin{equation}
\frac{\partial\vec{a}}{\partial u_i}\cdot\vec{\xi}_p=0 \Rightarrow
-\frac{1}{m}\frac{\partial V_i(x)}{\partial x} v_p=0.
\label{AlgebraicCond}
\end{equation}

The boundary condition for $\vec{\xi}_p$ is derived by variation
with respect to the final state:
\begin{equation}
 \left. \frac{\partial h}{\partial \vec{\xi}}\right|_{t_f}=0.
\end{equation}

\begin{figure}[t,b]
\centering
\begin{minipage}{0.6\linewidth}
\begin{center}
\includegraphics[width=.95\textwidth]{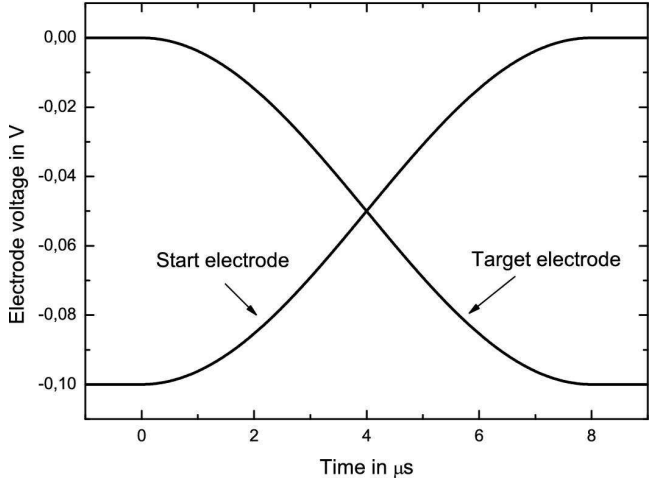}
\end{center}
\end{minipage}
\begin{minipage}{0.39\linewidth}
\caption{Initial guess for the control voltages, corresponding to
Eq. (\ref{InitialGuess}). \label{InitialGuess}}
\end{minipage}
\end{figure}

If we let the ion start at rest in the potential well pertaining to
the first electrode, the set of boundary conditions for the state
and costate vector reads
\begin{eqnarray}
x(t_0)=0 \nonumber \\
v(t_0)=0 \nonumber \\
x_p(t_f)=2\;(x-x_f) \nonumber \\
v_p(t_f)=2\;v. \label{BoundaryCond}
\end{eqnarray}

Eqs. (\ref{EqOfMotion}), (\ref{CoEqOfMotion}) and
(\ref{BoundaryCond}) together with (\ref{AlgebraicCond}) represent a
system of coupled ordinary nonlinear differential equations with
split boundary conditions, i.e. for two of the variables, initial
conditions are given whereas for the other two, the values at the
final time are specified. This makes a straightforward numerical
integration impossible. The system is therefore solved in an
iterative manner by means of a gradient search method. The scheme of
this steepest descent algorithm is as follows:

\begin{enumerate}
\item
Choose an initial guess for the control field $u_i(t)$.
\item Propagate $x$ and $v$ from $t=t_0$ to $t=t_f$ while using $u_i(t)$ in the corresponding equations of motion. At each time step, save the value of $x(t)$.
\item Determine $x_p(t_f)$ and $v_p(t_f)$ according to (\ref{BoundaryCond}).
\item Propagate $x_p$ and $v_p$ backwards in time from $t=t_f$ to $t=t_0$.
At each time step, save the value of $v_p(t)$.
\item For each time step, update the control field according to \begin{equation}
u_i^{new}(t)=u_i^{old}(t)+\tau\;v_p\;\frac{1}{m}\frac{\partial
V_1(x)}{\partial x} \label{FieldUpdate}
\end{equation}
\item Repeat steps 2 to 5 until the specified threshold fidelity is reached.
\end{enumerate}
In Eq.~(\ref{FieldUpdate}), the gradient search step width $\tau$ is
simply chosen by trial and error. If it is too small, the algorithm
converges too slowly, if it is too large, the algorithm starts to
oscillate. The values of $\alpha$ and $\beta$ in
Eq.~(\ref{CostAtFinalTime}) are determined based on experience. For
the data presented in the following section, these values are
$\alpha=10, \hspace{2mm} \beta=1$ and $\tau=5~\cdot~10^{-8}$. The
algorithm converged after about 200 iterations.

\begin{figure}[t,b]
\centering
\begin{minipage}{0.6\linewidth}
\begin{center}
\includegraphics[width=.95\textwidth]{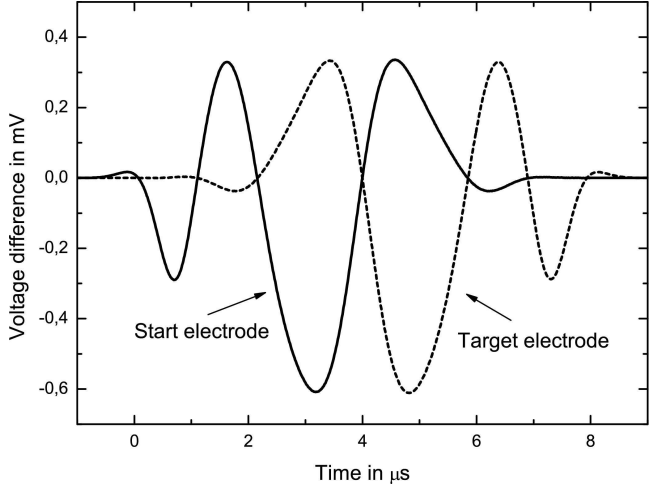}
\end{center}
\end{minipage}
\begin{minipage}{0.39\linewidth}
\caption{Optimized control voltages given in terms of change with
respect to the initial guess values. The solid curve indicates the
voltage $u_1(t)$ at the start electrode, the dashed one the voltage
$u_2(t)$ at the destination electrode. \label{Deltau}}
\end{minipage}
\end{figure}

\subsection{Optimization results}
\label{sec:results} For the initial guess the control field is
chosen as follows:
\begin{eqnarray}
u_0^{(0)}(t) &=&  \begin{cases} V_0 & \;\mbox{for}\; t\leq0 \cr
V_0\;\sin^2(\frac{\pi t}{2\Delta t}) &  \; \mbox{for}\;
0<t\leq\Delta t \cr 0 & \;\mbox{for}\; t>\Delta t
\end{cases}  \nonumber \\
u_1^{(1)}(t)& =& V_0-u_0^{(0)}(t) \label{InitialGuess}
\end{eqnarray}
This provides on the one hand a smooth and symmetric acceleration
and deceleration of the ion, on the other hand the potential minimum
exactly coincides with the desired positions
 at the initial and final times. In principle, other initial guess voltages like Gaussian
 flanks can be used as well. The reference voltage is $V_0=-0.1V$, corresponding to
 $\omega \approx 2\pi\cdot0.5\;$MHz in the initial and final potential wells. The switching
 time is set to $\Delta t=8.0\mu s$, the total time interval runs from $-1.0\mu s$
 to $9.0\mu s$.

\begin{figure}[t,b]
\begin{minipage}{0.6\linewidth}
\begin{center}
\includegraphics[width=0.95\textwidth]{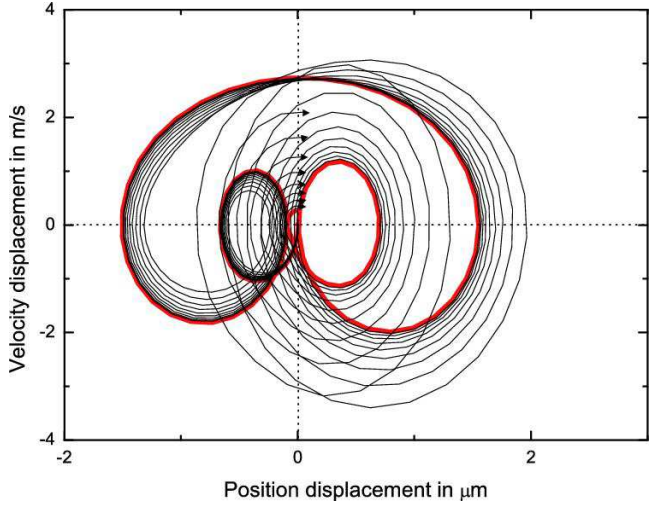}
\end{center}
\end{minipage}
\begin{minipage}{0.40\linewidth}
\caption{Phase space trajectories in the frame co-moving with the
potential minimum. The figure shows the trajectories pertaining to
iterations 0, 10 , ..., 100. At iteration 100, the ion arrives close
to the origin. Note that the trajectory tends to be symmetrized by
the optimization algorithm. \label{trajectory1}}
\end{minipage}
\end{figure}

\subsection{Ion heating due to anharmonic dispersion}
\label{anharmonic dispersion} Quantum mechanically we describe the
system with a Hamiltonian operator pertaining to a time-dependent
harmonic oscillator with an anharmonic perturbation:
\begin{equation}
H_0(t)=\frac{\hat{p}^2}{2m}+\frac{m\;\omega(t)^2}{2}(\hat{x}-x_0(t))^2+\kappa(t)(\hat{x}-x_0)^4.
\end{equation}
Without temporal variation of $\omega$ and the anharmonic
part\footnote{In contrast to $d_4$, $\kappa(t)$ is given by
expanding the potential around the instantaneous potential minimum.}
of the potential, the solution of the time-dependent Schr\"odinger
equation is simply given by a coherent state $|\alpha(t)\rangle$,
where the displacement parameter $\alpha(t)$ can be inferred from
the classical trajectory. Anharmonic dispersion of a wavepacket
occurs at a timescale given by $T_{rev}/(\Delta n)^2$
\cite{TANNOR2006}, with the revival time
\begin{equation}
T_{rev}=2h\left(\frac{d^2E_n}{dn^2}\right)^{-1}
\end{equation}
and the spread over the vibrational levels $\Delta n=\alpha(t)$. The
shift of the energy levels $E_n$ induced by the anharmonic
contribution causes a finite dispersion time and can be calculated
in first order stationary perturbation theory:
\begin{equation}
\Delta E_n(t)=\frac{5}{4}\frac{\hbar^2\kappa(t)}{m^2\omega(t)^2}n^2.
\end{equation}
We now define a generalized dispersion parameter
\begin{equation}
\int_{t_0}^{t_f}dt\frac{\Delta
n^2}{T_{rev}}=\frac{5\;\hbar}{4\pi\;m^2}\int_{t_0}^{t_f}dt
\frac{\kappa(t)|\alpha(t)|^2}{\omega(t)^2}.
\label{AnharmLossParam}
\end{equation}
If this parameter is sufficiently small, anharmonic dispersion will
not contribute to heating.

\begin{figure}[b]
\centering
\begin{center}
\includegraphics[width=.8\textwidth]{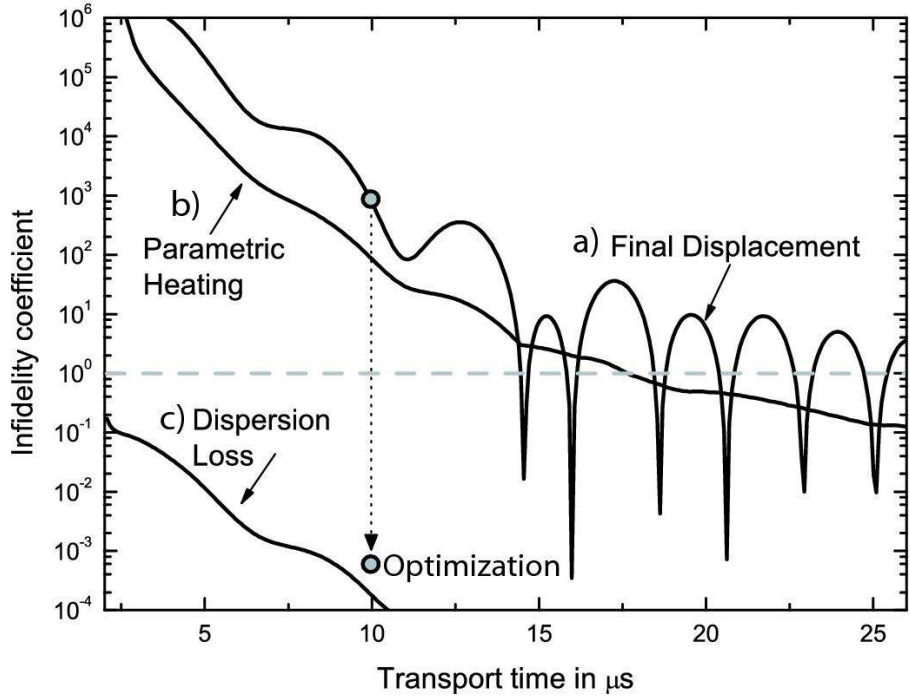}
\end{center}
\caption{Non-optimized transport, heating effects, and optimization
result: Indicated is the excess energy as a function to transport
time, if the initial guess function is used for the transport. The
curve (a) describes the energy at the final time in vibrational
quanta. The approximate zeros occur when the deceleration of the
potential minimum coincides with the ion oscillation. Curve (b)
displays the maximum non-adiabaticity parameter
Eq.~(\ref{ParamHeatingParameter}) for parametric heating and the (c)
the anharmonic dispersion loss parameter from
Eq.~(\ref{AnharmLossParam}). The grey dashed horizontal line
indicates a heating of one phonon and the borderline to
non-adiabatic behavior in curve (b). The optimization algorithm
decreases the excess displacement by more than 6 orders of
magnitude. The circles indicate the squared phase space displacement
for the guess function and the final optimized control
voltages.\label{Timescan}}
\end{figure}

\subsection{Quantum mechanical estimate of non-adiabatic parametric heating}
\label{parametric heating} We now check if the width of the
wavepacket adiabatically follows the harmonic frequency $\omega(t)$.
The adiabatic theorem states that if
\begin{equation}
\hbar\langle \phi_m(t)|\dot{\phi}_n(t)\rangle \ll |E_n(t)-E_m(t)|
\label{AdiabaticTheorem}
\end{equation}
is fulfilled, transitions between eigenstates can be neglected.
The parametric time dependence of the eigenstates states in Eq.
\ref{AdiabaticTheorem} is the implicit time dependence via
$\omega(t)$. We find the following nonvanishing matrix elements:
\begin{eqnarray}
\langle\phi_{n+1}|\phi_n\rangle &=&
\frac{\dot{\omega}}{\sqrt{2\pi^3}\omega}n\sqrt{n+1} \nonumber \\
\langle\phi_{n+2}|\phi_n\rangle &=&
\frac{\dot{\omega}}{4\omega}\sqrt{(n+1)(n+2)}
\label{ParamHeatingMatrixElem}
\end{eqnarray}
and similar expressions for $m=n-1,n-2$. Thus, parametric heating
can be neglected if
\begin{equation}
n^{3/2}\;\frac{\dot{\omega}}{\omega^2} \ll 1.
\label{ParamHeatingParameter}
\end{equation}
Numerical evaluation of the matrix elements yields the result that
the adiabatic following condition is fulfilled for $n=0$, but is
clearly violated for the high $n$ occurring for large excursion of
the wavepacket, for example $\bar{n}\approx 2000$ for $\Delta
x=1~\mu$m at a transport time of 10~$\mu$s, see Fig.~\ref{Timescan}.

\subsection{Improved initial guess function and ultra-fast transport}
We therefore have to refine our optimization strategy: As can be
seen in Fig.~(\ref{Deltau}), the control voltages changes are
symmetric, which indicates that one control degree of freedom can
be sacrificed in order to keep $\omega(t)$ constant. This is
achieved as follows: The initial guess voltages
Eq.~(\ref{InitialGuess}) are normalized to a constant $\omega$
before the optimization. The optimization process then leads to
variations in $\omega(t)$ that are negligibly small - typically
leading to maximum values of $\dot{\omega}/\omega^2$ on the order
of $10^{-5}$ such that according to
Eq.(\ref{ParamHeatingParameter}) the adiabatic theorem is
fulfilled even after the optimization algorithm has cured the
classical phase space displacement heating. This is in strong
contrast to the unconstrained, previous guess function, where we
obtain $\dot{\omega}/\omega^2\simeq 10^{-2}$. It should be noted
that parametric heating can be completely suppressed as well for
optimized control voltages. This can be achieved by changing the
set of control parameters to $\tilde{u}_1=u_1+u_2$ and
$\tilde{u}_2=u_1/\tilde{u}_1$. The new parameter $\tilde{u}_2$ is
now directly related to the instantaneous potential minimum $x_0$.
If only $\tilde{u}_2$ is incorporated in the optimization process,
$\tilde{u}_1$ can be readjusted at each step to keep $\omega$
constant.

The optimization results for the improved initial guess voltages
are shown in Fig.~\ref{TimeScan2}. The transport time could now be
reduced to 5$\mu$s which corresponds to roughly two oscillation
periods. For the improved guess funtion the wavepacket dispersion appears
now as the dominant heating source. This process
could be suppressed either by further geometric optimization of
the trap segments or by including a corresponding additional term
into the cost function of the optimization routine.

\begin{figure}[t,b]
\begin{minipage}{0.6\linewidth}
\begin{center}
\includegraphics[width=0.95\textwidth]{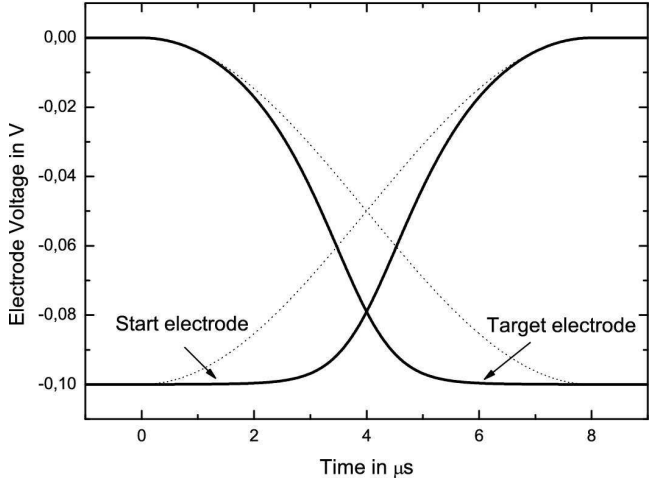}
\end{center}
\end{minipage}
\begin{minipage}{0.39\linewidth}
\caption{Initial guess voltages normalized to keep trap frequency
constant. The old initial guess voltages are indicated as dashed
curves. Note that the dynamics of the potential minimum is
unaffected by the normalization.\label{uInitialGuessImroved}}
\end{minipage}
\end{figure}

\subsection{Discussion of the open-loop result}

Our optimization results indicate that unwanted heating during ion
transport can be suppressed by many orders of magnitude by the
application of appropriate time-dependent control voltages.
Technically, one would achieve this using a fast high-resolution
digital-to-analog converter (DAC) with subsequent scaling to the
required voltage range. The small correction voltages obtained from
the optimization algorithm might represent a problem, however a 16
bit DAC with an appropriate scaling circuit would allow for a
discretization step of roughly 1.5$\mu$V for a maximum voltage
change of 0.1V. We have also checked the robustness of the control
field solutions against noise by calculating the trajectories with
white noise of variable level added on the voltages. We found a
quadratic dependence of the excess displacement on the noise level.
The deviation of the final displacement from the noise-free case was
negligibly small at a noise level of 20$\mu$V. Experimental values
for non-adiabatic heating effects in ion transport are given in Ref.
\cite{ROWE2002}. The comparison with our theoretical values is
hampered by the fact that these measurements have carried out at
higher axial trap frequency and the lighter ion species $^9$Be$^+$,
but over a much longer transport distance of $1.2$~mm. However, low
heating rates were obtained in those experiments only if the
transport duration corresponds to a relatively large number of about
$\simeq$ 100 of trap periods, whereas in our case, a transport
within only {\em two trap periods} was simulated.

\begin{figure}[t]
\begin{center}
\includegraphics[width=0.8\textwidth]{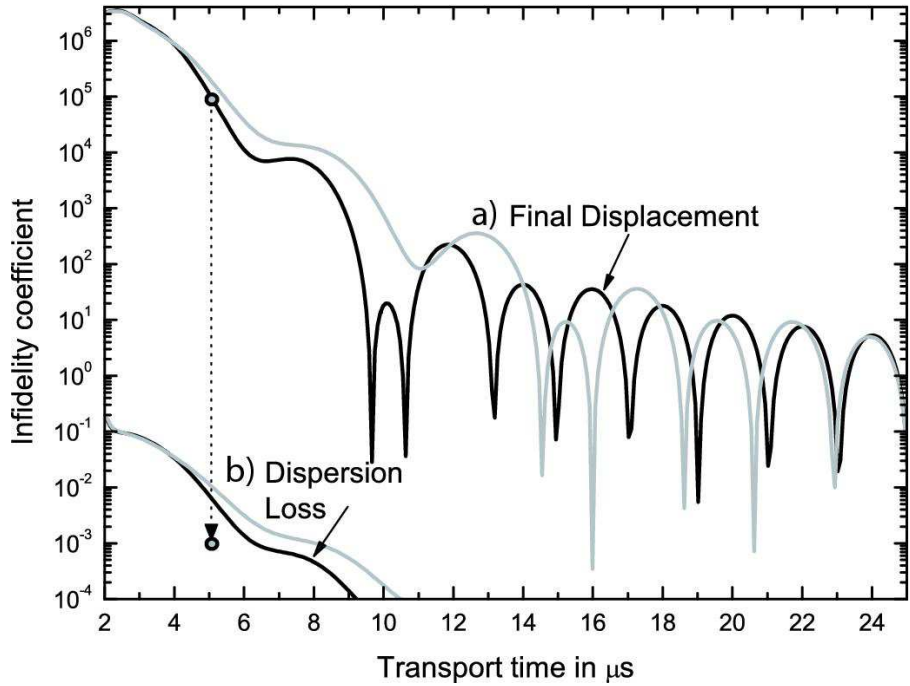}
\end{center}
\caption{Nonadiabatic effects versus transport time for the improved
initial guess voltages. Here, only the (a) excess displacement and
the (b) dispersion parameter are shown, parametric heating is not
relevant anymore. The values for the old initial guess voltages are
indicated in grey. The improved initial guess allows for successfull
optimization at a transport time of 5$\mu$s and an optimization of
about eight orders of magnitude in classical phase space
displacement. Now, with a few $\mu$s transport, anharmonic
dispersion becomes the predominant heating source.\label{TimeScan2}}
\end{figure}

\begin{figure}[t,b]
\begin{minipage}{0.6\linewidth}
\begin{center}
\includegraphics[width=0.95\textwidth]{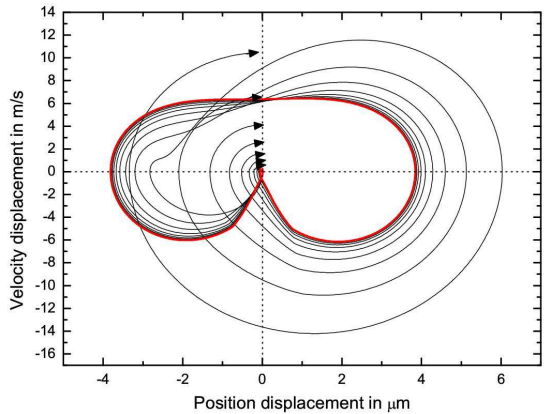}
\end{center}
\end{minipage}
\begin{minipage}{0.4\linewidth}
\caption{Phase space trajectories in the frame co-moving with the
potential minimum for the improved initial guess voltages. The
optimization is now carried out for transport time of only 5$\mu$s
corresponding to about two oscillation periods in the harmonic
trap potential. The thin lines indicate the optimization progress
and the fat line shows the final result after 100 iterations.
Again, the optimization routine symmetrized the trajectory.
\label{trajectory2}}
\end{minipage}
\end{figure}

\section{Outlook}
The optimization of ion transport beyond the speed limits given by
the anharmonic terms of the axial trapping potentials and parametric
heating would be most efficient and accurate if a full quantum
mechanical equation of motion was employed. Quantum mechanical
optimal control methods are based on the same variational principle
as presented here for a classical problem, with the only difference
that the terms in the penalty functional are functionals on Hilbert
space. Algorithms for quantum mechanical optimal control are well
developed and were applied to variety of different problems
\cite{CALARCO2004,SKLARZ2002}. In our case however, the application
of quantum mechanical optimal control was not yet possible for
simply a technical reason: The iterative solution via repeated
solution of Schr\"odinger equation over distances on the order of
200$\mu$m and time spans on the order of 20$\mu$s takes too much
computational effort, even with highly efficient methods like the
Fourier Grid Hamiltonian combined with the Chebyshev propagator
technique \cite{KosloffTimeDependentMethods}. On the other hand, we
have seen that for the typical electric potentials of segmented Paul
traps, the possibility to exert \textit{quantum} control on the
system is very restricted since the wavefunction of the ion mainly
senses a harmonic potential. The classical approach is therefore
well suited to the problem.

In future work, we will investigate whether quantum control could
be exerted during short time spans when the ion is displaced from
the potential minimum and therefore senses anharmonic
contributions to the potential. Extended Gaussian wave packet
dynamics \cite{GARRAWAY2000} could be used to take anharmonic
terms efficiently into account. Thus, the application of quantum
mechanical optimal control methods also opens new possibilities,
for example the control voltages could be used to devise new
schemes for quantum computational gates. In this case, the target
wave function for the optimization routine could be the first
excited motional state or even a superposition of different
motional Fock states. To fulfill this aim, anharmonic
contributions to the trapping potentials are crucial.

Open loop optimal control methods will also be applied to the
splitting of two ions \cite{POSCHINGER2006}. With this problem the
benefit of going beyond the adiabatic limit will be even more
promising. In an adiabatic manner the splitting is initiated by
lowering the steepness of the potential in order to increase the
separation of the two ions due to their mutual repulsion. This
decreases the trapping frequency and as a consequence the speed of
the procedure has to be decreased in order to stay adiabatic.

Open loop optimal control has proved to be successful for the
optimization of short broadband RF pulses in NMR experiments
\cite{SKINNER2005}. In a similar manner in ion trap based quantum
computing tailored, light pulses can speed up and improve
manipulation of the ions \cite{GARCIARIPOLL2003,RANGAN2004}. In
cases where analytical solution to the control problem is not
available open loop optimal control methods could be applied to
get optimized light pulses or electrostatic field configurations
for multi ion gate operations and entangled state preparation.

Another promising strategy that could be employed to avoid heating
during ion transport is the closed loop control technique. Here,
the experimental results are fed back into {\em e.g.} an
evolutionary algorithm to obtain improved values of the control
parameters. The heating rate can be measured by comparing the
strengths of the red and blue motional sidebands after the
transport process \cite{roos99}. The key problem for applying
closed loop control to ion transport lies in finding an
appropriate parametrization of the control voltages in order to
keep the parameter space small.

This technique may be applied equally well to the problem of
separation of two ions from one common potential into two
independent sections of the linear trap.

This work has merely started to apply the optimal control theory
for ion trap based quantum computing. Not only the motion of ions
between trap segments, but the entire process including shaped
laser pulses \cite{GARCIARIPOLL2003} and motional quantum state
engineering might be improved with this technique.

\begin{acknowledgement}
We acknowledge support by the European commission, the Deutsche
Forschungsgemeinschaft and by the Landesstiftung Baden-W\"urttemberg
within the frameworks "quantum highway A8" and "atomics". We thank
D. J. Tannor and T. Calarco for stimulating discussions.
\end{acknowledgement}

\begin{appendix}
\section{Comparison of our Boundary-Element-package with commercial software}
\label{Anhang BEM}

\begin{figure}[t]
\centering
\begin{center}
\includegraphics[width=1\textwidth]{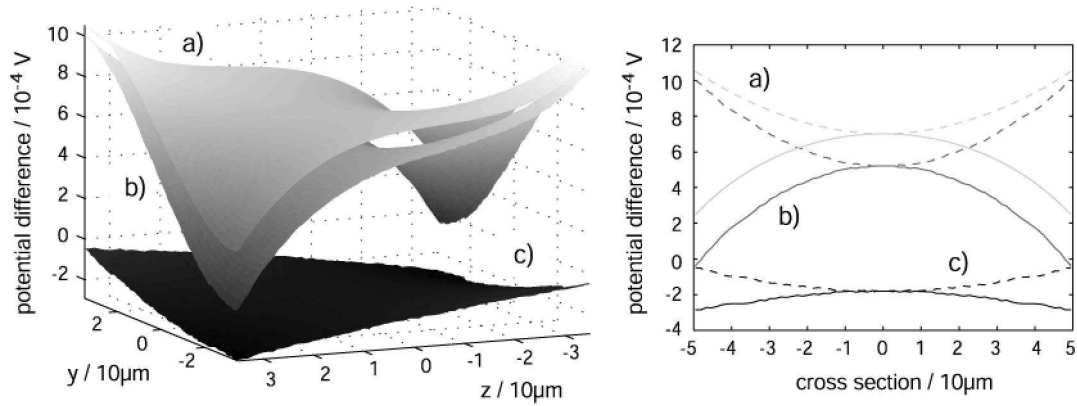}
\end{center}
\small \caption{a) Numeric result obtained by FEMLAB substracted
from the result of BEM \cite{download}, b) FEMLAB - CPO, c) CPO -
BEM \cite{download}. The left graph shows a 2D potential surface
plot in the y-z plane. The right graph shows a line plot in the
same plane in the direction of the two diagonal directions.}
\label{fig:BEM}
\end{figure}

Accurate values of the electrostatic potentials are of paramount
importance for the determination of the harmonic and the anharmonic
terms of the trapping potentials. An adequate choice of a numerical
solving method is the Boundary Element Method
(BEM)~\cite{POZRIKIDIS2002,BRKIC06}. BEM is a fast and more accurate
method compared to the Finite Element Method (FEM) or Finite
Difference Method (FDM). This is due to the fact that BEM only needs
to solve for the surface charges on the electrode surfaces. With
FEM/FDM the Laplace equation has to be solved on a three dimensional
mesh. Comparison of the speed and accuracy can be found in
\cite{CUBRIC1999}. In order to simplify the variation and
optimization of the trap geometry we have implemented a free
scriptable object oriented BEM package for 3D and 2D
\cite{download}. We have verified the results for the geometry of
Fig.~\ref{fig:3} of our package against the results of the
commercial BEM program CPO \cite{cpo} (see Fig.~\ref{fig:BEM}~c) and
against the results of the commercial FEM program FEMLAB
\cite{femlab} (see Fig.~\ref{fig:BEM}~a). The higher values in the
latter case are due to the inaccuracy of the finite element method
itself.


\end{appendix}


\end{document}